\documentclass[12pt]{article}
\usepackage{graphicx,amssymb,amsfonts,amsmath,epsfig,color}

\makeatletter
\DeclareSymbolFont{AMSb}{U}{msb}{m}{n}
\DeclareSymbolFontAlphabet{\mathbb}{AMSb}
\renewcommand{\section}{\@startsection{section}{1}{\z@}%
                                    {-7ex \@plus -1ex \@minus -.2ex}%
                                    {2.5ex \@plus.2ex}%
                                    {\normalfont\large\scshape\centering}}
\renewcommand{\subsection}{\@startsection{subsection}{2}{\z@}%
                                       {-5ex \@plus -1ex \@minus -.2ex}%
                                       {1.5ex \@plus.2ex}%
                                       {\normalfont\normalsize\scshape}}
\renewcommand{\subsubsection}{\@startsection{subsubsection}{3}{\z@}%
                                       {-5ex \@plus -1ex \@minus -.2ex}%
                                       {1.5ex \@plus.2ex}%
                                       {\normalfont\normalsize\scshape}}

\renewcommand\@seccntformat[1]{\ignorespaces\csname #1name\endcsname\space
                               \csname the#1\endcsname.\quad}   % Extra period and name added
\newdimen\captionmargin
\newdimen\captionindent
\newdimen\captionwidth
\newcommand{\captionfont}{\slshape}
\newcommand\@captionlabel[1]{\textsc{#1:}\space}
\long\def\@makecaption#1#2{%
  \vskip\abovecaptionskip
  \captionwidth\hsize
  \advance\captionwidth -2\captionmargin
  \sbox\@tempboxa{\@captionlabel{#1}\captionfont #2}%
  \ifdim \wd\@tempboxa >\captionwidth
    \ifdim\captionindent>\z@
      \advance\captionwidth -\captionindent
      \hskip\captionindent
    \fi
    \hskip\captionmargin
    \parbox[t]{\captionwidth}{\leavevmode\hskip-\captionindent
      \@captionlabel{#1}\captionfont #2}%
  \else
    \global \@minipagefalse
    \hb@xt@\hsize{\hfil\box\@tempboxa\hfil}%
  \fi
  \vskip\belowcaptionskip}
\def\eqnarray{%
   \stepcounter{equation}%
   \def\@currentlabel{\p@equation\theequation}%
   \global\@eqnswtrue
   \m@th
   \global\@eqcnt\z@
   \tabskip\@centering
   \let\\\@eqncr
   $$\everycr{}\halign to\displaywidth\bgroup
       \hskip\@centering$\displaystyle\tabskip\z@skip{##}$\@eqnsel
      &\global\@eqcnt\@ne$\;\hfil{##}$\hfil
      &\global\@eqcnt\tw@$\;\displaystyle{##}$\hfil\tabskip\@centering
      &\global\@eqcnt\thr@@ \hb@xt@\z@\bgroup\hss##\egroup
         \tabskip\z@skip
      \cr}
\ifcase \@ptsize
\or
\or
\fi
  \divide\@tempdima\baselineskip
  \@tempcnta=\@tempdima
\makeatother
\begin{document}

%
%--------- Titlepage ---------------------------------------------------------
%
% useful for drafts %%%%%%%%%%%%%%%%%%%%%%%%%%%%%%%%%%%%%%%%%%%%%%%%%%%%%%%%
%\pagestyle{myheadings} \markboth{PRELIMINARY VERSION}{PRELIMINARY
%VERSION 1}
%\newcommand\query[1]
%{\marginpar{\begin{flushleft}\footnotesize#1\end{flushleft}}}
%%%%%%%%%%%%%%%%%%%%%%%%%%%%%%%%%%%%%%%%%%%%%%%%%%%%%%%%%%%%%%%%%%%%%%%%%%%%
% --- This bit puts ``draft'' over everything! ---
%\special{!userdict begin /bop-hook{gsave 200 30 translate
%65 rotate /Times-Roman findfont 216 scalefont setfont
%0 0 moveto 0.93 setgray (DRAFT) show grestore}def end}
%  --- end of this bit that puts `draft' over everything ---

\renewcommand{\theequation}{\arabic{section}.\arabic{equation}}
\renewcommand{\thefigure}{\arabic{figure}}
\newcommand{\gapprox}{%
\mathrel{%
\setbox0=\hbox{$>$}\raise0.6ex\copy0\kern-\wd0\lower0.65ex\hbox{$\sim$}}}
\textwidth 165mm \textheight 220mm \topmargin 0pt \oddsidemargin 2mm
\def\ib{{\bar \imath}}
\def\jb{{\bar \jmath}}

\newcommand{\ft}[2]{{\textstyle\frac{#1}{#2}}}
\newcommand{\be}{\begin{equation}}
\newcommand{\ee}{\end{equation}}
\newcommand{\bea}{\begin{eqnarray}}
\newcommand{\eea}{\end{eqnarray}}
\newcommand{\Identity}{{1\!\rm l}}% Unit Matrix
\newcommand{\cx}{\overset{\circ}{x}_2}
\def\CN{$\mathcal{N}$}
\def\CH{$\mathcal{H}$}
\def\hg{\hat{g}}
\newcommand{\bref}[1]{(\ref{#1})}
\def\espai{\;\;\;\;\;\;}
\def\zespai{\;\;\;\;}
\def\avall{\vspace{0.5cm}}
\newtheorem{theorem}{Theorem}
\newtheorem{acknowledgement}{Acknowledgment}
\newtheorem{algorithm}{Algorithm}
\newtheorem{axiom}{Axiom}
\newtheorem{case}{Case}
\newtheorem{claim}{Claim}
\newtheorem{conclusion}{Conclusion}
\newtheorem{condition}{Condition}
\newtheorem{conjecture}{Conjecture}
\newtheorem{corollary}{Corollary}
\newtheorem{criterion}{Criterion}
\newtheorem{defi}{Definition}
\newtheorem{example}{Example}
\newtheorem{exercise}{Exercise}
\newtheorem{lemma}{Lemma}
\newtheorem{notation}{Notation}
\newtheorem{problem}{Problem}
\newtheorem{prop}{Proposition}
\newtheorem{rem}{{\it Remark}}
\newtheorem{solution}{Solution}
\newtheorem{summary}{Summary}
\numberwithin{equation}{section}
\newenvironment{pf}[1][Proof]{\noindent{\it {#1.}} }{\ \rule{0.5em}{0.5em}}
\newenvironment{ex}[1][Example]{\noindent{\it {#1.}}}

\thispagestyle{empty}

%\begin{flushright}\scshape
%January 2004
%\end{flushright}
%\vskip1cm

\begin{center}

{\LARGE\scshape Some considerations about NS5 and LST Hawking radiation
\par}
\vskip15mm

\textsc{Oscar Lorente-Esp\'{i}n}
\par\bigskip
%$^a${\em
%Departament d'Estructura i Constituents de la Mat\`eria,
%Universitat de Barcelona,\\
%Diagonal 647, E-08028 Barcelona, Spain.}\\[.1cm]
{\em
Departament de F{\'\i}sica i Enginyeria Nuclear,
Universitat Polit\`ecnica de Catalunya,\\
Comte Urgell, 187, E-08036 Barcelona, Spain.}\\[.1cm]
\vspace{5mm}
\end{center}

\section*{Abstract}
We have studied the Hawking radiation corresponding to the NS5 and Little String Theory (LST) black hole models using two 
semi-classical methods: the complex path method and a gravitational anomaly.
After summarizing some known concepts about the thermodynamics of these theories, we have computed the emission rates for 
the two black hole models. The temperature calculated from, e.g. the well-known surface gravity expression, is shown to be 
identical to that obtained from both the computation of the gravitational anomaly and the complex path method. 
Moreover, the two semi-classical methods show that NS5 exhibits non-thermal behavior that contrasts with the thermal 
behavior of LST. We remark that energy conservation is the key factor leading to a non-thermal profile for NS5. In contrast, 
LST keeps a thermal profile even when energy conservation is considered because temperature in this model does not depend on 
energy.

\vskip10mm
\noindent KEYWORDS: Black Holes, String Theory, Hawking radiation.

\vspace{3mm} \vfill{ \hrule width 5.cm \vskip 2.mm {\small
\noindent E-mail: oscar.lorente-espin@upc.edu }}

%%%% Old version
%We have studied the Hawking radiation of the NS5 and Little String Theory (LST) using complex path method and gravitational
%anomalies. After some comments about the thermodynamics of the theories, we have computed the emission rates of the two 
%models. Next, we have find agreement between the temperature obtained in the computation of the gravitational anomaly and the 
%temperature obtained by other methods, like surface gravity. Both semi-classical methods show the non-thermal behavior of 
%NS5 in contrast with the thermal behavior of LST. We point out the importance of the energy conservation in order to get a 
%non thermal spectrum. LST, even though the non energy dependence of the temperature, keeps the thermal profile. In the 
%discussion, we have verified that cluster decomposition does not work for NS5, however it works for LST. Afterwards we 
%make a brief digression about non-local aspects and cluster decomposition in LST. Finally, we have computed 
%the temperature measured by an inertial particle-like observer at fixed distance $r$ away from the event horizon of NS5 and 
%LST. We can check that the Hagedorn temperature is effectively the maximum temperature. 

%\vspace{3mm} \vfill{ \hrule width 5.cm \vskip 2.mm {\small
%\noindent E-mail: pons@ecm.ub.es, pere.talavera@upc.es }}

\newpage
\setcounter{page}{1}

%%%%%%%%%%%%%%%%%%%%%%%%%%%%%%%%%%%%%%%%%%%%%%%%%%%%%%%%%%%%%%%%
%%%%%%%%%%%%%%%%%%%%%%%%%%%%%%%%%%%%%%%%%%%%%%%%%%%%%%%%%%%%%%%%%%

%%%%%%%%%%%%%%%%%%%%%%%%
\tableofcontents       %
\vskip 1cm             %
%%%%%%%%%%%%%%%%%%%%%%%%

\setcounter{equation}{0}

\section{Introduction}
Since the pioneering proposal of Hawking that black holes can radiate 
\cite{Hawking:1974sw}, much work has been done in order to obtain a complete theory of quantum gravity. When 
Hawking announced his amazing results, a new powerful paradox emerged. The information loss paradox with the apparent 
violation of unitarity principle has consequences on well-established quantum mechanics. A recent effort in order to solve 
this 
paradox has been done studying different semi-classical approaches such as the tunneling method proposed by 
Parikh and Wilczek 
\cite{Parikh:1999mf, Parikh:2004ih}, the complex path analysis \cite{Srinivasan:1998ty, Shankaranarayanan:2000gb, 
Shankaranarayanan:2000qv} or the cancellation of gravitational anomalies \cite{Robinson:2005pd, Iso:2006wa, Banerjee:2008az}.
\par
%{Iso:2006ut, Banerjee:2007qs, Banerjee:2007uc, Banerjee:2008az, Banerjee:2008wq, Gangopadhyay:2008ub}.
We have studied the Hawking radiation for NS5 and LST stringy black holes using different semi-classical methods obtaining 
equal results for each method. For good reviews on LST and NS5, we address the readers to 
\cite{Kutasov:2001uf, Aharony:1999ks, Kapustin:1999ci, Kutasov:2000jp, Rangamani:2001ir, Aharony:1998ub}. 
We have verified 
that the NS5 model shows a non-thermal emission whereas LST shows a thermal emission. This last conclusion matches with the 
Hagedorn properties of LST, namely the temperature of LST corresponds to the Hagedorn temperature.\par
The Letter is organized as follows: in section 2, we briefly summarize the LST theory with some properties and thermodynamics. 
In section 3, we reduce the ten-dimensional metric of LST to two-dimensional one. All the physics will be analyzed within
the propagation of massless particles in the $r-t$ sector of the metric. 
In section 4, we study two different semi-classical methods in order to compute the emission rate for NS5 and LST. Complex 
path method and anomalies yields the same results as the tunneling method, analyzed in \cite{LorenteEspin:2007gz}, 
for the temperature and the emission rate. It is worth to mentioning that in the classical computation of the Bogoliubov 
coefficients all the results for emission rates shows thermal profiles due to the lack of energy conservation. 
This fact had driven Hawking to state that all the 
information that falls into the black hole is lost for ever, establishing in this way the information loss paradox. 
Nevertheless, one hopes to overcome this weird conclusion using semi-classical methods.   

\section{A glance at LST thermodynamics}
Little string theory is a non-gravitational six-dimensional and non-local field theory 
\cite{Kutasov:2001uf, Aharony:1999ks, Kapustin:1999ci, Peet:1998wn, Seiberg:1997zk}, believed 
to be dual to a string theory background, defined as the decoupled theory on a stack of $N$ NS5-branes. In the 
limit of a vanishing asymptotic value for the string coupling $g_{s}\rightarrow0$, keeping the string length $l_{s}$ 
fixed while the energy above extremality is fixed, i.e. $\frac{E}{m_{s}}={\rm fixed}$, the processes in which the 
modes that live on 
the branes are emitted into the bulk as closed strings are suppressed. The theory becomes free in the bulk, but 
strongly interacting on the brane. In this limit, the theory reduces to Little String Theory or more precisely to 
(2,0) LST for type IIA NS5-branes and to (1,1) LST for type IIB NS5-branes \cite{Aharony:1998ub}.\par
The throat geometry corresponding to $N$ coincident non-extremal NS5-branes in the string frame \cite{Maldacena:1997cg} is

\begin{equation}
 \label{metric}
ds^{2}=-f(r)dt^2+\frac{A(r)}{f(r)}dr^2+A(r)r^2d\Omega_{3}^2+\sum_{j=1}^{5}dx_{j}^2\;,
\end{equation}
where $dx_{j}^2$ corresponds to flat spatial directions along the 5-branes, $d\Omega_{3}^2$ corresponds to 3-sphere of
the transverse geometry and the dilaton field is defined as $e^{2\Phi}=g_{s}^{2}A(r)$. The metric functions are defined as

\begin{equation}
 \label{metricf}
f(r)=1-\frac{r_{0}^2}{r^2}\;\;\;,\;\;\;A(r)=\chi+\frac{N}{m_{s}^2r^2}\;,
\end{equation}
$r_{0}$ is the non-extremality parameter, so the extremal configuration is obtained by the limit $r_{0}\rightarrow0$
and the location of the event horizon corresponds to $r=r_{0}$. We define the parameter $\chi$ which takes the values 
1 for NS5 model and 0 for LST, these are only the values for which exist a supergravity solution. In addition to the 
previous fields there 
is an $NS-NS$ $H_{(3)}$ form along the $S^3$, $H_{(3)}=2N\epsilon_{3}$. According to the holographic principle the high
spectrum of this dual string theory should be approximated by certain black hole in the background (\ref{metric}). 
The geometry transverse to the 5-branes is a long
tube which opens up into the asymptotic flat space with the horizon at the other end, in the limit $r\rightarrow r_{0}$
appears the semi-infinite throat parametrized by $(t,r)$ coordinates, in this region the dilaton grows linearly pointing out 
that gravity becomes strongly coupled far down the throat. The string propagation in this 
geometry should correspond to an exact conformal field theory \cite{Kutasov:1990ua}. The boundary of the near horizon 
geometry is $R^{5,1}\times R\times S^{3}$. The geometry (\ref{metric}) is regular as long as $r_{0}\neq 0$. In order to 
construct the thermal states of the black holes 
we write the corresponding metric in Rindler coordinates as
\begin{equation}
 \label{rindler}
ds^{2}=-\kappa^2\rho^{2}dt^2+d\rho^{2}+A(r)r^2d\Omega_{3}^2+\sum_{j=1}^{5}dx_{j}^2\;,
\end{equation}
where we have introduced the radial Rindler coordinate $\rho$ (proper length), also we see that the quantity
$\kappa=\frac{f(r_{0})'}{2\sqrt{A(r_{0})}}$ coincides with the surface gravity of the NS5 and LST black holes. 
Then performing a Wick rotation $t_{E}=it$, the 
Euclidean time coordinate has to be periodic with period $2\pi$ in order to avoid a conical singularity. Thus
we identify the period of the Euclidean time as in \cite{Cotrone:2007qa}
\begin{equation}
 \label{beta}
\beta=\frac{2\pi}{\kappa}=\frac{2\pi\sqrt{N+\chi r_{0}^2}}{m_{s}}\;.
\end{equation}
The temperature obtained, $T=\beta ^{-1}$, does not depend on which frame we are using, the string frame and 
Einstein frame are related by a local rescaling that does not affect the result.\par  
Regarding LST model, we notice that the temperature is independent of the black hole radius and therefore of the black 
hole mass. 
%This leads to a complete degenerate
%thermodynamical phase space \cite{Maldacena:1996ya}, in this work the calculation of Bekenstein-Hawking entropy 
%suggests a microscopic description for a gas of strings living on a near extremal 5-brane yielding a density of states of 
%Hagedorn type . 
In this way we could identify this temperature with the Hagedorn temperature of the superstring theory.
One can compute the energy density for the LST background in ten dimensions, $e\equiv\frac{E}{V_{5}}$ and the 
entropy density, 
$s\equiv\frac{S}{V_{5}}$, where $V_{5}$ is the volume of the flat 5-branes space and $S$ is the standard Bekenstein-Hawking
entropy calculated from the area of the event horizon of the black hole, $S=\frac{\rm Area}{4G_{10}}$. Either in Einstein 
frame metric \cite{Cotrone:2007qa, Harmark:2000hw} or in string frame metric \cite{Rangamani:2001ir, Maldacena:1997cg} it 
is satisfied the usual thermodynamic relation $S=\beta E$. This relation implies that the free energy 
of the system $F=E-TS$ vanishes. At very high energies the equation of state is of the Hagedorn form which leads to an 
exponentially growing density of states \cite{Kutasov:2000jp}: $\rho(E)=e^{S(E)}\sim e^{\beta E}$.\par
At first sight one could think that a phase transition is present when the system evolves from NS5 to the near
horizon limit of NS5, i.e. LST, but we have checked that it is not the case. Computing the Bekenstein-Hawking entropy

\begin{equation}
 \label{entropy}
S=\frac{\rm Area}{4G_{10}}=\frac{V_{5}}{2G_{10}}\pi^2(N+\chi m_{s}^2r_{0}^2)^{3/2}\;,
\end{equation}
and plotting it versus the temperature we do not detect any critical point (Davies point) \cite{La:2010bx} 
that would signal a phase transition. Even working in thermodynamic geometry \cite{Banerjee:2010bx}, writing the LST
metric like a Ruppeiner metric $ds^{2}=-3\sqrt{\frac{\pi G}{\hbar^{2}M}}dS^2$, we do not detect any divergence in the scalar 
curvature that would signal a possible phase
transition. However calculating the specific heat as $C=T\frac{\partial S}{\partial T}$, we have found that it has
a negative value, $-3S$, showing that the theory is unstable. In the work \cite{Kutasov:2000jp},
the authors show that loop/string corrections to the Hagedorn density of states of LST would be of the form 
$\rho(E)\sim E^{\alpha}e^{\beta E}(1+O(\frac{1}{E}))$ and the temperature-energy relation becomes 
$\beta=\frac{\partial \log\rho}{\partial E}=\beta_{0}+\frac{\alpha}{E}+O(\frac{1}{E^2})$. They found that since $\alpha$ 
is negative the high energy thermodynamics corresponding to near-extremal 5-branes is unstable as well as the temperature 
is above the Hagedorn temperature and the specific heat is negative. This instability would be associated
to the presence of a negative mode (tachyon) in string theory, the high temperature phase of the theory yields the 
condensation of this mode. The authors are lead again to the conclusion that the Hagedorn temperature is reached at a 
finite energy being associated with a phase transition.\par
Next we would like to address the question whether an observer in a moving frame observes a temperature above the 
Hagedorn temperature.
We know that in the near horizon limit of NS5, i.e. LST, the system reaches the maximum temperature, namely the Hagedorn 
temperature. One could think that a boosted observer may observes a temperature bigger than the Hagedorn one, for this 
reason we want to verify the validity of this statement.
We have evaluated the simplest case, a scalar particle-like observer which moves on an NS5-brane with constant velocity at 
fixed distance $r$ from the horizon of the LST black hole. We consider the orbit for which $y_{1}=vt$. 
Relating the time coordinate $t$ with the proper time $\tau$ through $d\tau^2=-g_{\mu\nu}dx^{\mu}dx^{\nu}$, 
one obtains

\begin{equation}
 \label{propertime}
\frac{d\tau}{dt}=\sqrt{f(r)-v^2} \;.
\end{equation}
The velocity is bounded by the local velocity of light thus we have to impose the constraint $v^2\leq f(r)$. 
This relation brings us to a new coordinate of the horizon position, where causality is lost, seen by the moving particle,
$r=\frac{r_{0}}{\sqrt{1-v^2}}$. Furthermore 
the Killing vector relevant for the process is $\zeta=-\partial_{t}+v\partial_{y_{1}}$. 
Therefore evaluating the surface gravity at this new coordinate $r$, we obtain the local temperature for the 
moving scalar particle

\begin{equation}
 \label{tempinertial}
T^{'}=\frac{(1-v^2)\;m_{s}}{2\pi\sqrt{N+\chi\frac{r_{0}^2}{(1-v^2)}}}\;,
\end{equation}
where we have worked out in natural units, $c=1$ and $v<1$. We notice two important features. First of all, we see that in 
the 
$v\rightarrow0$ limit we recover the result (\ref{beta}). Secondly, comparing the temperature for the particle-like observer
(\ref{tempinertial}) with the temperature defined by (\ref{beta}) for an asymptotic static observer, we see that the 
former is lower than the second one. We conclude that the Hawking temperature of LST is a maximum bound and corresponds 
to the Hagedorn temperature. Unfortunately, we are not able to perform the same analysis for an accelerating particle-like 
observer. The main problem is that the path which the particle follows is not generated by a Killing vector field, this 
fact prevent us from using the surface gravity method in order to calculate the temperature.
  
\section{Effective two-dimensional theory}
We are going to reduce the metric (\ref{metric}) to the $r-t$ sector, relevant for the forthcoming 
sections. At first step we take the scalar field action

\begin{equation}
 \label{action}
S=\frac{1}{2\kappa_{10}^2}\int_{M}d^{10}x\sqrt{-g}\left(R-\frac{1}{2}\partial_{\mu}\phi\partial^{\mu}\phi
-\frac{1}{12}e^{-\Phi}H_{(3)}^{2}\right)\;\;.
\end{equation}
Performing a change to tortoise coordinate, see (\ref{metric}) $dr^{*}=\frac{\sqrt{A(r)}}{f(r)}dr$, 
we expand the ten-dimensional action as 

%\begin{eqnarray*}
% \label{actiontortoise}
%\lefteqn{S=} \nonumber \\
%& &\frac{1}{2\kappa_{10}^2}\int dtdr^{*}d\theta d\varphi d\psi \prod_{j=2}^{6}dx_{j}\;r^{3}A(r)^{2}sin^{2}\theta\;
%sin\varphi\;e^{-2\Phi}\left[\frac{f(r)}{\sqrt{A(r)}}\left(R -\frac{e^{-\Phi}}{12}H_{(3)}^{2}\right)+\nonumber\\
%& &\left(\frac{1}{2\sqrt{A(r)}}(\partial_{t}^{2}-\partial_{r^{*}}^{2})-
%\frac{f(r)}{2r^{2}A^{3/2}}\left(\partial_{\theta}^{2}+
%\frac{1}{sin^{2}\theta}\partial_{\varphi}^{2} +\frac{1}{sin^{2}\theta sin^{2}\varphi}\partial_{\psi}^{2}\right)-
%\frac{f(r)}{2\sqrt{A(r)}}\sum_{j=2}^{6}\partial_{j}^{2}\right)\nonumber\\
%\lefteqn{\phi(t,r)S(\Omega_{3})\;e^{i\sum k_{j}x_{j}}\right]}
%\end{eqnarray*}

%\begin{align}
% \label{actiontortoise1}
%S&= \frac{1}{2\kappa_{10}^2}\int dtdr^{*}d\theta d\varphi d\psi \prod_{j=2}^{6}dx_{j}\;r^{3}A(r)^{2}sin^{2}\theta\;
%sin\varphi\;e^{-2\Phi}\left[\frac{f(r)}{\sqrt{A(r)}}\left(R -\frac{e^{-\Phi}}{12}H_{(3)}^{2}\right) +\nonumber\\
%&\left(\frac{1}{2\sqrt{A(r)}}(\partial_{t}^{2}-\partial_{r^{*}}^{2})-\frac{f(r)}{2r^{2}A^{3/2}}\left(\partial_{\theta}^{2}
%+\frac{1}{sin^{2}\theta}\partial_{\varphi}^{2} +\frac{1}{sin^{2}\theta sin^{2}\varphi}\partial_{\psi}^{2}\right)-
%\frac{f(r)}{2\sqrt{A(r)}}\sum_{j=2}^{6}\partial_{j}^{2}\right) \nonumber\\
%&\phi(t,r)S(\Omega_{3})\;e^{i\sum k_{j}x_{j}} \right] \\
%\end{align}

\begin{equation}
 \label{actiontortoise1}
\begin{split}
S&= \frac{1}{2\kappa_{10}^2}\int dtdr^{*}d\theta d\varphi d\psi \prod_{j=1}^{5}dx_{j}\;r^{3}A(r)^{2}sin^{2}\theta\;
sin\varphi\;(g_{s}e^{-\Phi})^{5/2}\left[\frac{f(r)}{\sqrt{A(r)}}\left(R -\frac{e^{-\Phi}}{12}H_{(3)}^{2}\right) +\right.\\
&\left(\frac{1}{2\sqrt{A(r)}}(\partial_{t}^{2}-\partial_{r^{*}}^{2})-\frac{f(r)}{2r^{2}A^{3/2}}\left(\partial_{\theta}^{2}
+\frac{1}{sin^{2}\theta}\partial_{\varphi}^{2} +\frac{1}{sin^{2}\theta sin^{2}\varphi}\partial_{\psi}^{2}\right)-
\frac{f(r)}{2\sqrt{A(r)}}\sum_{j=2}^{6}\partial_{j}^{2}\right)\\
&\left. \phi(t,r)S(\Omega_{3})\;e^{i\sum k_{j}x_{j}} \bigg] \;,
\right.\end{split} 
\end{equation}
where we have decomposed the scalar field into $r-t$, 3-angular and 5-brane parts.
Our following approximations are based on three main steps:
\begin{enumerate}
\item We only consider the propagation mode of an s-wave.
\item We only take into account a subset of states of the Hilbert space such that the eigenstates of momentum 
parallel to the NS5-brane vanish.
\item We take the near horizon limit, $r\rightarrow r_{0}$.
\end{enumerate}
 \par

%The key point is to establish the near horizon limit for which the metric function $f(r)$ vanish. Near horizon we 
%also have considered the propagation of an s-wave neglecting so the angular terms. 
%If we take a subset of states of the Hilbert space so that the eigenstates of momentum parallel to the 5-brane are zero
%we can neglect the propagation on the brane.
Eventually we come back to the original $r$ radial coordinate obtaining for the action  

\begin{equation}
 \label{action2d}
S=\frac{V(S^3)V_{5}}{2\kappa_{10}^2}\int dtdr A(r)^2 e^{-2\Phi}\; 
\left(-\frac{1}{f(r)}\partial_{t}^2+\frac{f(r)}{A(r)}\partial_{r}^2\right)\phi(t,r) \;,
\end{equation}
where $V(S^3)$ is the volume of the 3-sphere and $V_{5}$ the volume of 5-branes. From (\ref{action2d}) we find out 
that the scalar field can be seen as $(1+1)$-dimensional scalar field $\phi(t,r)$ propagating in the background

\begin{equation}
 \label{effmetric}
ds_{eff}^2=-f(r)dt^2+\frac{A(r)}{f(r)}dr^2 \;,
\end{equation}
together with an effective dilaton field

\begin{equation}
 \label{dilatonfield}
e^{2\Phi}=g_{s}^{2}A(r) \;.
\end{equation} 
Henceforth we are going to work with this two-dimensional effective metric.

\section{Semi-classical Black Hole emission}
\subsection{Complex path method}
In this section we are going to show a semi-classical method to obtain the Hawking radiation. Eventually we will obtain 
the same 
thermal behavior than in \cite{Hawking:1974sw}, where the particle production is computed using the Bogoliubov 
transformation.
The complex path method has been developed in \cite{Srinivasan:1998ty}, in order to calculate particle production in 
Schwarzschild-like space-time and it was extended for a different coordinate representations of the Schwarzschild space-time 
\cite{Shankaranarayanan:2000gb, Shankaranarayanan:2000qv}. Nevertheless complex path analysis had already been discussed by 
Landau \cite{Landau:1975gn}, where it was used to describe tunneling processes in non-relativistic semi-classical quantum 
mechanics.\par
We will follow the reference \cite{Srinivasan:1998ty} where the authors avoid to work in Kruskal representation. They
use the standard coordinates in the $r-t$ sector. However the method presents a disadvantage because one would find a 
coordinate singularity at the horizon. Nevertheless using the techniques of complex integration one bypasses 
the singularity.
We also want to mention that the method of complex path leads to the same results that in 
\cite{Hartle:1976tp}. In both methods, for the Schwarzschild space-time and also as we will see in this section 
for NS5 and LST space-time, it has been found that the relation between emission and absorption probabilities is of the form

\begin{equation}
\label{relprob1}
P_{e}=e^{-\beta\omega}P_{a} \;,
\end{equation}
where $\omega$ is the energy of the emitted particles. We are tempting to compare this relation with the standard thermal 
Boltzmann distribution for blackbody radiation where $\beta^{-1}$ is identified with Hawking temperature. We have verified 
that this is the case, if we compare our results versus the temperature calculated using the 
definition of surface gravity for example. It is noteworthy to say that this method allows 
one to get temperatures for black holes only comparing probabilities of emission and absorption but is not able to 
calculate the spectrum of thermal radiation. In that sense the tunneling method 
proposed is so far incomplete. To amend this shorts the authors in \cite{Banerjee:2009wb} present a new mechanism. \\
In order to apply the complex path method to NS5 and LST we have constructed the semi-classical action obtained from 
Hamilton-Jacobi equations. Afterwards we have computed the semi-classical propagator $K(r_{2},t_{2};r_{1},t_{1})$. 
Eventually we have calculated the emission and absorption probabilities.\par
We consider the equation of motion of a massless scalar particle $\square\phi=0$, in the background (\ref{effmetric})

\begin{equation}
\label{hjeom1}
-A(r)\frac{\partial^2}{\partial t^2}\phi(t,r)+
\frac{f(r)}{r^3}\frac{\partial}{\partial r}\left[r^3f(r)\frac{\partial}{\partial r}\phi(t,r)\right]=0 \;.
\end{equation}
Using the standard ansatz solution 
%ref: manual cuantica (feynmann, sakurai,...)
\begin{equation}
\label{hjansatz}
\phi(t,r)\sim e^{\frac{i}{\hbar}S(t,r)} \;,
\end{equation}
and substituting in (\ref{hjeom1}) we get an expression in terms of the action $S(t,r)$ 

\begin{equation}
\label{hjeom2}
-A(r)\left(\frac{\partial S}{\partial t}\right)^2+f(r)^2\left(\frac{\partial S}{\partial r}\right)^2+
\frac{\hbar}{i}\left[-A(r)\frac{\partial^2 S}{\partial t^2}+f(r)^2\frac{\partial^2 S}{\partial r^2}+
\frac{f(r)}{r^3}\frac{d(r^3f(r))}{dr}\frac{\partial S}{\partial r}\right]=0\;,
\end{equation}
where we have collected the terms with $\hbar$ dependence. The following step is to write the action as an 
expansion in a power series of $(\frac{\hbar}{i})$

\begin{equation}
\label{hjactionexp}
S(t,r)=S_{0}(t,r)+\left(\frac{\hbar}{i}\right)S_{1}(t,r)+\left(\frac{\hbar}{i}\right)^2S_{2}(t,r)+... \;.
\end{equation}
Substituting the above expansion in (\ref{hjeom2}) and neglecting terms of the order $(\frac{\hbar}{i})$ and 
higher, we obtain a non-linear first order partial differential equation which corresponds to the Hamilton-Jacobi 
equation of motion to the leading order in the action $S$,

\begin{equation}
\label{hjeom3}
-A(r)\left(\frac{\partial S_{0}(t,r)}{\partial t}\right)^2+
f(r)^{2}\left(\frac{\partial S_{0}(t,r)}{\partial r}\right)^2=0\;.
\end{equation}
We are interested in the evaluation of the semi-classical propagator which inform us about the  amplitude 
for a particle going from $r_{1}$ at time $t_{1}$ to $r_{2}$ at time $t_{2}$. In the saddle point approximation we get

\begin{equation} 
\label{hjk1}
K(r_{2},t_{2};r_{1},t_{1})=N\exp\left[\frac{i}{\hbar}S_{0}(r_{2},t_{2};r_{1},t_{1})\right] \;,
\end{equation}
where $N$ is a normalization constant. From (\ref{hjeom3}) we get

\begin{equation}
\label{hjs1}
S_{0}(r_{2},t_{2};r_{1},t_{1})=-\omega(t_{2}-t_{1})\pm\omega\int_{r_{1}}^{r_{2}}\frac{\sqrt{A(r)}}{f(r)}dr \;,
\end{equation}
the plus/minus sign correspond to ingoing/outgoing particles respectively and $\omega$ is the energy of the emitted or 
absorbed particles.\par
The integral (\ref{hjs1}) is not well behaved if the horizon $r_{0}$ is within the region of integration. 
However this turns to be the case, we are interested in the emission of particles through the event horizon, 
so the region of 
integration runs from inside the horizon to outside. \par
First we consider the propagation of an 
outgoing particle in the inner region $r_{1}<r_{0}$. Applying the usual complex analysis tools, we deform the contour of 
integration around the pole $r_{0}$ in the upper complex half-plane. Obtaining for the radial part of (\ref{hjs1})
\begin{equation}
\label{hjs6}
S_{0}^{e}=\frac{i\pi\omega}{2}r_{0}\sqrt{A(r_{0})}\;.
\end{equation}
We will call it emission action because we simply consider the emission of an outgoing particle propagating from inside 
the horizon to the outside.\par
By the same talk one proceeds with analogous analysis to evaluate the action at lowest 
order for absorbed particles. In that case we are considering the propagation of an ingoing particle in the outer region, 
$r_{0}<r_{2}$. Deforming the contour of integration in the upper complex half-plane, eventually we obtain the same result 
as the emission process up to change of sign. Now we are obtaining the absorption action for a particle that propagates 
from the region outside of the horizon to the inside 
\begin{equation}
\label{hjs9}
S_{0}^{a}=-\frac{i\pi\omega}{2}r_{0}\sqrt{A(r_{0})}\;.
\end{equation}
We are interested in the expressions (\ref{hjs6}) and (\ref{hjs9}) in order to evaluate the probabilities of the emission 
and absorption processes. Thereby using the definition of the probability: $P=|K(r_{2},t_{2};r_{1},t_{1})|^2$,
and substituting the expression for the corresponding actions, we finally obtain for the emission and absorption 
probabilities

\begin{equation}
\label{hjprobem}
P_{e}\sim \exp\left[-\frac{\pi}{\hbar}\omega r_{0}\sqrt{A(r_{0})}\right]\;,\;
P_{a}\sim \exp\left[\frac{\pi}{\hbar}\omega r_{0}\sqrt{A(r_{0})}\right] \;,
\end{equation}
where we have omitted the normalization constants. Eventually we are interested in to write the relation between emission 
and absorption probabilities

\begin{equation}
\label{hjprobrel}
P_{e}=\exp\left[-\frac{2\pi}{\hbar}\omega r_{0}\sqrt{A(r_{0})}\right]P_{a}\;.
\end{equation}
At first sight we observe that absorption process dominate over the emission, it is more easy for the system to absorb 
than to radiate particles. Also we note some misleading form in the expression for the absorption probability 
(\ref{hjprobem}) because we could think that one might get a probability absorption greater than $1$. Nevertheless we 
only have considered the spatial contribution of 
the action in order to calculate the probabilities of emission and absorption processes. Instead of this we should also 
have considered the time contribution as proposed in the work \cite{deGill:2010nb}.\par
Comparing (\ref{hjprobrel}) with the same relation in a thermal bath of particles (\ref{relprob1}) we can identify the 
temperature of our system (taking $\hbar=1$ and $m_{s}=1$)

\begin{equation}
\label{hjtemp1}
T=\frac{1}{2\pi r_{0}\sqrt{A(r_{0})}}=\frac{1}{2\pi\sqrt{\chi r_{0}^2+N}} \;,
\end{equation}
that coincides with the value of temperature obtained in 
(\ref{beta}).\par
So far we have studied NS5/LST systems without backreaction. The next step is to consider the backreaction of the metric 
due to the emission process. For all the details we address the reader to \cite{LorenteEspin:2007gz} where we had studied 
the tunneling method. Generically the process consists in the emission of a particle with 
energy $\omega$ from a black hole to the background. Taking into account the energy conservation the metric backreacts. 
Hence the total ADM mass is conserved and consequently the black hole mass must decrease by the same amount of the energy 
that it has been released. Our starting point in the evaluation of the backreaction is the expression of the action for the 
emission process (\ref{hjs6}). In our NS5/LST model we have the following relation between the event horizon and the mass 
of the black hole: $r_{0}^2\sim M$, where $M$ is the mass of the black hole and 
the factors missed are not relevant in our study. When the metric backreacts in the emission process the energy 
conservation implies that $r_{0}^2\rightarrow r_{0}^2-\omega$. The shrink of the event horizon rides the tunneling 
emission between turning points defined just inside and just outside of the event horizon. Once the emission has been 
carried out we can perform the previous change in (\ref{hjs6})

\begin{equation}
\label{hjsback1}
S_{0}^{e}=\frac{i\pi}{2}\omega\sqrt{\chi(r_{0}^2-\omega)+N}\;,
\end{equation}
expanding for low energies we get

\begin{equation}
\label{hjsback2}
S_{0}^{e}=\frac{i\pi}{2}\left(\omega\sqrt{\chi r_{0}^2+N}-\frac{\chi\omega^{2}}{2\sqrt{\chi r_{0}^2+N}}+
O(\omega)^{3}\right)\;.
\end{equation} 
Calculating the emission probability for both models we obtain
%\begin{equation}
%\label{hjprobembackns5}
%P_{e}\sim \exp\left[-\frac{\pi}{\hbar}\left(\omega\sqrt{r_{0}^2+N}-\frac{\omega^{2}}{2\sqrt{r_{0}^2+N}}+
%...\right)\right]\;.
%\end{equation}
%\begin{equation}
%\label{hjprobembacklst}
%P_{e}\sim \exp\left[-\frac{\pi}{\hbar}\omega\sqrt{N}\right]\;.
%\end{equation}
%\begin{equation}
% \label{hjprobemback}
%P_{e}=\begin{cases}\begin{matrix} \exp\left[-\frac{\pi}{\hbar}\left(\omega\sqrt{r_{0}^2+N}-
% \frac{\omega^{2}}{2\sqrt{r_{0}^2+N}}+...\right)\right] & if\;\chi=1 \\
% \exp\left[-\frac{\pi}{\hbar}\omega\sqrt{N}\right] & if\;\chi=0
%\end{matrix}\end{cases} \;.
%\end{equation}

\begin{equation}
\label{hjprobemback}
P_{e}\sim\left\{
\begin{array}{ccc}
\exp\left[-\frac{\pi}{\hbar}\left(\omega\sqrt{r_{0}^2+N}-\frac{\omega^{2}}{2\sqrt{r_{0}^2+N}}+
...\right)\right] & {\rm if\;\;\chi=1\;\;(NS5)}; \cr
  &  & \cr
\exp\left[-\frac{\pi}{\hbar}\omega\sqrt{N}\right] & {\rm if\;\;\chi=0\;\;(LST).}
\end{array} 
\right. \;
\end{equation}
We see higher order correction terms corresponding to the NS5 emission probability, which indicates that the 
emission is not purely thermal. On the other hand the emission probability expression corresponding to the LST model 
is exact, which indicates that the emission is purely thermal.\par
As a final comment we want to stress the crucial fact of the energy conservation in order to get a non-thermal emission in 
the tunneling formalism. In this work, we have concluded that the results obtained from the tunneling formalism in 
\cite{Parikh:1999mf} are nothing more than an extension of the Hamilton-Jacobi formalism taking into account the energy 
conservation, which induces the backreaction of the event horizon. In our particular case we are facing with an 
anomalous model which does not accomplish the previous expectations about non-thermal emission. The LST model emits 
thermal radiation in any case with or without energy conservation.\par
In order to analyze the deviations from the thermal behavior of the NS5 model it would be relevant to perform 
the computation of the greybody factors. Thus we must solve the radial part of the equation of motion (\ref{hjeom1}). 
As far we know this equation cannot be solved analytically, therefore it is necessary a numerical analysis which 
enlighten the non-thermal aspects of the NS5 model. Even so, we can elucidate that the non-thermal behavior of the NS5 
model comes from the throat region. In this region the dilaton grows linearly pointing out that gravity becomes strongly 
coupled far down the throat, and states with large quantum numbers exist. On the other hand the near horizon limit of 
the NS5, i.e. LST, decouples the mode interactions between the bulk and the brane, the spectrum reduces to 
less excited states, leading to a thermal behavior with the Hagedorn temperature, see \cite{Madrigal:2009nf} for a 
complete discussion.

\subsection{Anomalies}
In this section we want to present another successful semi-classical method to compute the Hawking radiation from an 
evaporating black hole. The method is based on the cancellation of gravitational anomalies in a two-dimensional chiral 
theory taken as effective theory near the event horizon. This method was first proposed in 
\cite{Robinson:2005pd}. Gravitational anomalies are anomalies in general covariance, i.e. general coordinate 
transformations (diffeomorphism), and they manifests as the non-conservation of the energy-momentum tensor.\par
The authors in \cite{Robinson:2005pd, Iso:2006wa} 
managed the treatment of gauge and covariant anomalies deriving an effective two-dimensional theory close 
to the horizon. They built an effective action performing a partial wave decomposition in tortoise coordinate and dropping 
potential factors which vanish exponentially fast near the horizon. Thus physics near the horizon can be described by an 
infinite collection of (1+1) fields with the metric reduced to the $r-t$ sector. In these previous works the authors derived 
the Hawking radiation flux by anomaly cancellation, splitting the space-time into the near horizon region where the anomaly 
works and outside region when the conservation law is preserved. They carried out the calculation using the consistent 
chiral anomaly form of the energy-momentum tensor, see \cite{AlvarezGaume:1983ig, Bertlmann:2000da},

\begin{equation}
\label{angravcons}
\nabla_{\mu}T^{\mu}_{\nu}=
\frac{1}{96\pi\sqrt{-g}}\epsilon^{\beta\delta}\partial_{\delta}\partial_{\alpha}\Gamma^{\alpha}_{\nu\beta}
\end{equation} 
together with the covariant boundary condition at the horizon. It is the cancellation of the covariant anomaly which 
lead us to the appearance of the Hawking radiation flux.\par
On the other hand it is known the existence of two types of anomalies. Covariant anomalies, which transform covariantly 
under gauge or general coordinate transformation but they do not satisfy the Wess-Zumino consistency condition. 
And consistent anomalies, which satisfy the consistency condition but they do not transform covariantly under gauge or 
general coordinate transformation. In our study we adopt the procedure carried out in \cite{Banerjee:2008az} 
where the authors use a more coherent frame, working with covariant forms both for the expression of the chiral anomaly 
and for the boundary conditions. Unlike the previous works it is not necessary split the space-time into different regions, 
near the horizon region and outside. \par
First of all we consider the physics near the horizon of the NS5 and LST models described by an infinite collection of (1+1) 
scalar field particles propagating in the background (\ref{effmetric}).
It is not necessary to work with the full metric because only the $r-t$ sector is relevant for the emission processes, 
obtaining so the same results for the full theory as for the effective two-dimensional theory.
In this frame we can consider that only the outgoing modes are present. The ingoing 
modes are lost into the black hole and they do not affect at the classical level. Nevertheless the total effective action 
must be covariant. Thereby the quantum contribution of these irrelevant ingoing modes will supply the extra term, 
a Wess-Zumino term, 
in order to cancel the gravitational anomaly providing the Hawking flux \cite{Iso:2006wa}. The loss of the ingoing modes 
behind the horizon of the black hole causes that the effective theory becomes chiral, obtaining consequently a 
gravitational anomaly \cite{AlvarezGaume:1983ig, Bertlmann:2000da}. Following \cite{Banerjee:2008az}, we adopt the 
expression for the covariant form of the gravitational anomaly 

\begin{equation}
\label{angravcov}
\nabla_{\mu}T^{\mu\nu}=\frac{1}{96\pi\sqrt{-g}}\epsilon^{\nu\mu}\nabla_{\mu}R\;,
\end{equation} 
where $R$ is the Ricci scalar and $\epsilon^{\nu\mu}$ is the Levi-Civit\'{a} tensor that in our case takes the values 
$\epsilon^{tr}=-\epsilon^{rt}=1$ and zero for other contributions. The covariant boundary condition at the event horizon is

\begin{equation}
\label{anbccov}
T^{r}_{t}(r=r_{0})=0\,.
\end{equation}
Noticing that we are working with a static metric, we evaluate the equation (\ref{angravcov}) for the effective 
two-dimensional theory in the $r-t$ sector. Eventually we get

\begin{equation}
\label{angravcov1}
\partial_{r}(\sqrt{-g}T^{r}_{t})=\frac{1}{96\pi}g_{tt}\partial_{r}R \;.
\end{equation} 
The Ricci scalar for NS5 and LST models is

\begin{equation}
\label{anricci}
R=\frac{f'A'}{2A^2}-\frac{f''}{A}\;,
\end{equation}
where primes denotes derivative with respect to the coordinate $r$. Defining the new function

\begin{equation}
\label{anN}
N^{r}_{t}\equiv\frac{1}{96\pi}(-\frac{ff'A'}{2A^2}-\frac{f'^{2}}{2A}+\frac{ff''}{A})\;,
\end{equation}
we can write (\ref{angravcov1}) as

\begin{equation}
\label{angravcov2}
\partial_{r}(\sqrt{-g}T^{r}_{t})=\partial_{r}N^{r}_{t} \;.
\end{equation} 
Then integrating the equation (\ref{angravcov2}) we obtain

\begin{equation}
\label{anem1}
\sqrt{-g}T^{r}_{t}=b_{0}+(N^{r}_{t}(r)-N^{r}_{t}(r_{0}))\,,
\end{equation}
where $b_{0}$ is an integration constant that can be evaluated implementing the covariant boundary condition 
(\ref{anbccov}). Doing so it yields the value $b_{0}=0$. Hence (\ref{anem1}) becomes

\begin{equation}
\label{anem2}
T^{r}_{t}=\frac{1}{\sqrt{-g}}(N^{r}_{t}(r)-N^{r}_{t}(r_{0}))\;.
\end{equation}
The Hawking radiation flux is measured at infinity where the covariant gravitational anomaly vanishes. Therefore we 
compute the energy flux by taking the asymptotic limit of (\ref{anem2})

\begin{equation}
\label{anem3}
T^{r}_{t}(r\rightarrow\infty)=-\frac{1}{\sqrt{-g}}N^{r}_{t}(r_{0})\;.
\end{equation}
Evaluating (\ref{anN}) at the event horizon $r_{0}$ and considering the value of the surface gravity 
$\kappa=\frac{1}{\sqrt{N+\chi r_{0}^{2}}}$, we finally obtain for the energy flux at infinity

\begin{equation}
\label{anem4}
T^{r}_{t}(r\rightarrow\infty)=\frac{1}{\sqrt{-g}}\frac{\kappa^{2}}{48\pi}\;,
\end{equation}
which it is of course the Hawking radiation flux for a black hole.

\section{Conclusion-Discussion}

In this work we have started reviewing briefly some aspects about LST thermodynamics. We have exposed the thermal emission 
of LST due to the non-energy dependence of the Hagedorn temperature. Also we have evaluated the temperature experienced 
by a scalar particle-like observer, thereby we have verified that the Hagedorn temperature of LST is a 
maximum bound. Furthermore we have studied the Hawking radiation of the NS5 and LST black hole models using two 
semi-classical emission methods: the complex path method and the cancellation of the gravitational anomaly. We want to 
mention that using both methods we have recovered the previous results in \cite{LorenteEspin:2007gz} where we worked 
using the tunneling formalism. The complex path method 
\cite{Srinivasan:1998ty, Shankaranarayanan:2000qv} shows how to evaluate the emission rate in the framework of the 
Hamilton-Jacobi
formalism. We have shown that imposing energy conservation, in order to take into account the backreaction of the metric 
during the emission process, we reproduce exactly the same results that in the tunneling formalism proposed in 
\cite{Parikh:1999mf, Parikh:2004ih}.
We would like to point out the advantage that represents the complex path method with respect to the tunneling formalism. 
First of all, we avoid heuristic explanations about the tunneling mechanism in the process of the emission. 
Secondly, we work with the well-known Hamilton-Jacobi equations plus the imposition of the energy 
conservation. And finally, it is not necessary to change the standard coordinates of the metric into Painlev\'{e} 
coordinates. We 
conclude that the tunneling method is nothing more than the complex path method plus energy conservation. 
Nevertheless none of the above methods are able to clarify the information loss paradox since
they do not calculate the spectrum.\par
We have verified that another successful method to evaluate Hawking radiation in NS5 and LST models is based on the 
cancellation of the gravitational anomaly \cite{Robinson:2005pd}. \par 
We have shown that all the above methods lead to a non-thermal emission for the NS5 model and thermal emission for the 
LST model, see (\ref{hjprobemback}).\par 
Finally, cluster decomposition principle is a crucial physical requirement which states that very distant experiments 
produced uncorrelated results, thus establishing the local behavior of the field theory. Cluster decomposition principle 
states that if multi-particle processes are performed in $N$ very distant laboratories, then the S-matrix element for the 
overall process factorizes. This factorization ensures a factorization of the corresponding transition probabilities, 
corresponding to uncorrelated experimental results.
In the line of the works \cite{Zhang:2009jn, Zhang:2009td} where the authors linked the existence of correlations 
among tunneled particles and the entropy conservation of the full system (black hole plus Hawking radiation), 
we have calculated the successive emission probabilities for two particles  of energies $\omega_{1}$ and $\omega_{2}$ 
using (\ref{hjprobemback}) for each model respectively. We have found that the NS5 model does not satisfy cluster 
decomposition

\begin{equation}
 \label{ns5cluster}
\ln\mid\Gamma(\omega_{1}+\omega_{2})\mid-\ln\mid\Gamma(\omega_{1})\Gamma(\omega_{2})\mid=
\frac{\omega_{1}\omega_{2}}{2\sqrt{N+r_{0}^{2}}}\,.
\end{equation}
On the other hand we have found that the LST model satisfies cluster decomposition as we expected

\begin{equation}
 \label{lstcluster}
\ln\mid\Gamma(\omega_{1}+\omega_{2})\mid-\ln\mid\Gamma(\omega_{1})\Gamma(\omega_{2})\mid=0\,,
\end{equation}
where $\Gamma(\omega_{1})$ and $\Gamma(\omega_{2})$ are the emission probabilities corresponding to a particle of energy 
$\omega_{1}$ and $\omega_{2}$ respectively; $\Gamma(\omega_{1}+\omega_{2})$ is the emission probability of a particle with 
energy $\omega_{1}+\omega_{2}$. We have found that $\Gamma(\omega_{1},\omega_{2})=\Gamma(\omega_{1}+\omega_{2})$ is 
accomplished at low energies, namely, the emission probability of a particle $\omega_{2}$ conditioned by the previous 
emission of a particle $\omega_{1}$ is the same as the emission of a single particle of energy $\omega_{1}+\omega_{2}$.
With these results at hand we can conclude that in the NS5 black hole exists correlations between emitted particles. 
This fact is intimately related with the non-thermal emission rate (\ref{hjprobemback}). Regarding (\ref{ns5cluster})
one hopes that the successive Hawking emissions could preserve unitarity avoiding in such a way the information
loss paradox. However it is not the case for the LST black hole where the thermal emission rate 
(\ref{hjprobemback}) lead us to cluster decomposition. Therefore the successive emissions of particles are independent 
one of each other, thus the information of the initial states remain hidden.\par
%It has been demonstrated that LST is a non-local theory. LST exhibits T-duality therefore ensures that it is a non-local 
%theory, moreover LST shows an impossibility to be Fourier transformed into position space
%\cite{Kutasov:2001uf, Aharony:1999ks, Kapustin:1999ci, Peet:1998wn}. On the other hand LST exhibits cluster decomposition 
%pointing out a local behavior. In order to match the non-locality of LST with cluster decomposition property, one is lead 
%to the presence of Yukawa-like interactions mediated by non-massless particles within the theory.

\vskip10mm
\noindent{\bf Acknowledgments:}\\
I thank Pere Talavera for fruitful discussions and insightful comments on this work. Also I appreciate a lot the 
instructive comments formulated by F. J. Perez-Reche.


\begin{thebibliography}{33}

%\cite{Hawking:1974sw}
\bibitem{Hawking:1974sw}
  S.~W.~Hawking,
  ``Particle Creation by Black Holes,''
  Commun.\ Math.\ Phys.\  {\bf 43 } (1975)  199-220.

%\cite{Parikh:1999mf}
\bibitem{Parikh:1999mf}
  M.~K.~Parikh, F.~Wilczek,
  ``Hawking radiation as tunneling,''
  Phys.\ Rev.\ Lett.\  {\bf 85 } (2000)  5042-5045.
  [hep-th/9907001].

%\cite{Parikh:2004ih}
\bibitem{Parikh:2004ih}
  M.~K.~Parikh,
  ``A Secret tunnel through the horizon,''
  Int.\ J.\ Mod.\ Phys.\  {\bf D13 } (2004)  2351-2354.
  [hep-th/0405160].

%\cite{Srinivasan:1998ty}
\bibitem{Srinivasan:1998ty}
  K.~Srinivasan, T.~Padmanabhan,
  ``Particle production and complex path analysis,''
  Phys.\ Rev.\  {\bf D60 } (1999)  024007.
  [gr-qc/9812028].

%\cite{Shankaranarayanan:2000gb}
\bibitem{Shankaranarayanan:2000gb}
  S.~Shankaranarayanan, K.~Srinivasan, T.~Padmanabhan,
  ``Method of complex paths and general covariance of Hawking radiation,''
  Mod.\ Phys.\ Lett.\  {\bf A16 } (2001)  571-578.
  [gr-qc/0007022].

%\cite{Shankaranarayanan:2000qv}
\bibitem{Shankaranarayanan:2000qv}
  S.~Shankaranarayanan, T.~Padmanabhan, K.~Srinivasan,
  ``Hawking radiation in different coordinate settings: Complex paths approach,''
  Class.\ Quant.\ Grav.\  {\bf 19 } (2002)  2671-2688.
  [gr-qc/0010042].

%\cite{Robinson:2005pd}
\bibitem{Robinson:2005pd}
  S.~P.~Robinson, F.~Wilczek,
  ``A Relationship between Hawking radiation and gravitational anomalies,''
  Phys.\ Rev.\ Lett.\  {\bf 95 } (2005)  011303.
  [gr-qc/0502074].

%\cite{Iso:2006wa}
\bibitem{Iso:2006wa}
  S.~Iso, H.~Umetsu, F.~Wilczek,
  ``Hawking radiation from charged black holes via gauge and gravitational anomalies,''
  Phys.\ Rev.\ Lett.\  {\bf 96 } (2006)  151302.
  [hep-th/0602146].

%\cite{Banerjee:2008az}
\bibitem{Banerjee:2008az}
  R.~Banerjee,
  ``Covariant Anomalies, Horizons and Hawking Radiation,''
  Int.\ J.\ Mod.\ Phys.\  {\bf D17 } (2009)  2539-2542.
  [arXiv:0807.4637 [hep-th]].

%\cite{Kutasov:2001uf}
\bibitem{Kutasov:2001uf}
  D.~Kutasov,
  ``Introduction to little string theory,''
  Lectures given at the Spring School on Superstrings and Related Matters. Trieste, 2-10 April 2001.

%\cite{Aharony:1999ks}
\bibitem{Aharony:1999ks}
  O.~Aharony,
  ``A Brief review of 'little string theories',''
  Class.\ Quant.\ Grav.\  {\bf 17 } (2000)  929-938.
  [hep-th/9911147].

%\cite{Kapustin:1999ci}
\bibitem{Kapustin:1999ci}
  A.~Kapustin,
  ``On the universality class of little string theories,''
  Phys.\ Rev.\  {\bf D63 } (2001)  086005.
  [hep-th/9912044].

%\cite{Kutasov:2000jp}
\bibitem{Kutasov:2000jp}
  D.~Kutasov, D.~A.~Sahakyan,
  ``Comments on the thermodynamics of little string theory,''
  JHEP {\bf 0102 } (2001)  021.
  [hep-th/0012258].

%\cite{Rangamani:2001ir}
\bibitem{Rangamani:2001ir}
  M.~Rangamani,
  ``Little string thermodynamics,''
  JHEP {\bf 0106 } (2001)  042.
  [hep-th/0104125].

%\cite{Aharony:1998ub}
\bibitem{Aharony:1998ub}
  O.~Aharony, M.~Berkooz, D.~Kutasov {\it et al.},
  ``Linear dilatons, NS five-branes and holography,''
  JHEP {\bf 9810 } (1998)  004.
  [hep-th/9808149].

%\cite{LorenteEspin:2007gz}
\bibitem{LorenteEspin:2007gz}
  O.~Lorente-Espin, P.~Talavera,
  ``A Silence black hole: Hawking radiation at the Hagedorn temperature,''
  JHEP {\bf 0804 } (2008)  080.
  [arXiv:0710.3833 [hep-th]].

%\cite{Peet:1998wn}
\bibitem{Peet:1998wn}
  A.~W.~Peet, J.~Polchinski,
  ``UV / IR relations in AdS dynamics,''
  Phys.\ Rev.\  {\bf D59 } (1999)  065011.
  [hep-th/9809022].

%\cite{Seiberg:1997zk}
\bibitem{Seiberg:1997zk}
  N.~Seiberg,
  ``New theories in six-dimensions and matrix description of M theory on T**5 and T**5 / Z(2),''
  Phys.\ Lett.\  {\bf B408 } (1997)  98-104.
  [hep-th/9705221].

%\cite{Callan:1991at}
%\bibitem{Callan:1991at}
%  C.~G.~Callan, Jr., J.~A.~Harvey, A.~Strominger,
%  ``Supersymmetric string solitons,''
%  In *Trieste 1991, Proceedings, String theory and quantum gravity '91* 208-244 and Chicago Univ. - EFI 91-066 
%  (91/11,rec.Feb.92) 42 p.
%  [hep-th/9112030].

%\cite{Maldacena:1997cg}
\bibitem{Maldacena:1997cg}
  J.~M.~Maldacena, A.~Strominger,
  ``Semiclassical decay of near extremal five-branes,''
  JHEP {\bf 9712 } (1997)  008.
  [hep-th/9710014].

%\cite{Kutasov:1990ua}
\bibitem{Kutasov:1990ua}
  D.~Kutasov, N.~Seiberg,
  ``Noncritical superstrings,''
  Phys.\ Lett.\  {\bf B251 } (1990)  67-72.

%\cite{Cotrone:2007qa}
\bibitem{Cotrone:2007qa}
  A.~L.~Cotrone, J.~M.~Pons, P.~Talavera,
  ``Notes on a SQCD-like plasma dual and holographic renormalization,''
  JHEP {\bf 0711 } (2007)  034.
  [arXiv:0706.2766 [hep-th]].

%\cite{Harmark:2000hw}
\bibitem{Harmark:2000hw}
  T.~Harmark, N.~A.~Obers,
  ``Hagedorn behavior of little string theory from string corrections to NS5-branes,''
  Phys.\ Lett.\  {\bf B485 } (2000)  285-292.
  [hep-th/0005021].

%\cite{Maldacena:1996ya}
%\bibitem{Maldacena:1996ya}
%  J.~M.~Maldacena,
%  ``Statistical entropy of near extremal five-branes,''
%  Nucl.\ Phys.\  {\bf B477 } (1996)  168-174.
%  [hep-th/9605016].

%\cite{La:2010bx}
\bibitem{La:2010bx}
  H.~La,
  ``Davies Critical Point and Tunneling,''
  
  [arXiv:1010.3626 [hep-th]].

%\cite{Banerjee:2010bx}
\bibitem{Banerjee:2010bx}
  R.~Banerjee, S.~K.~Modak, S.~Samanta,
  ``A New Phase Transition and Thermodynamic Geometry of Kerr-AdS Black Hole,''
  
  [arXiv:1005.4832 [hep-th]].

%\cite{Landau:1975gn}
\bibitem{Landau:1975gn}
  L.~D.~Landau, E.~M.~Lifshitz,
  ``Textbook On Theoretical Physics. Vol. 2: Quantum Mechanics.''
  Pergamon Press, New York, (1975).

%\cite{Hartle:1976tp}
\bibitem{Hartle:1976tp}
  J.~B.~Hartle, S.~W.~Hawking,
  ``Path Integral Derivation of Black Hole Radiance,''
  Phys.\ Rev.\  {\bf D13 } (1976)  2188-2203.

%\cite{Banerjee:2009wb}
\bibitem{Banerjee:2009wb}
  R.~Banerjee, B.~R.~Majhi,
  ``Hawking black body spectrum from tunneling mechanism,''
  Phys.\ Lett.\  {\bf B675 } (2009)  243-245.
  [arXiv:0903.0250 [hep-th]].

%\cite{deGill:2010nb}
\bibitem{deGill:2010nb}
  A.~de Gill, D.~Singleton, V.~Akhmedova {\it et al.},
  ``A WKB-like approach to Unruh Radiation,''
  Am.\ J.\ Phys.\  {\bf 78 } (2010)  685-691.
  [arXiv:1001.4833 [gr-qc]].

%\cite{Madrigal:2009nf}
\bibitem{Madrigal:2009nf}
  J.~D.~Madrigal, P.~Talavera,
  ``A Note on the string spectrum at the Hagedorn temperature,''
  [arXiv:0905.3578 [hep-th]].


%\cite{Iso:2006ut}
%\bibitem{Iso:2006ut}
%  S.~Iso, H.~Umetsu, F.~Wilczek,
%  ``Anomalies, Hawking radiations and regularity in rotating black holes,''
%  Phys.\ Rev.\  {\bf D74 } (2006)  044017.
%  [hep-th/0606018].

%\cite{Banerjee:2007qs}
%\bibitem{Banerjee:2007qs}
%  R.~Banerjee, S.~Kulkarni,
%  ``Hawking radiation and covariant anomalies,''
%  Phys.\ Rev.\  {\bf D77 } (2008)  024018.
%  [arXiv:0707.2449 [hep-th]].

%\cite{Banerjee:2007uc}
%\bibitem{Banerjee:2007uc}
%  R.~Banerjee, S.~Kulkarni,
%  ``Hawking radiation, effective actions and covariant boundary conditions,''
%  Phys.\ Lett.\  {\bf B659 } (2008)  827-831.
%  [arXiv:0709.3916 [hep-th]].

%\cite{Banerjee:2008wq}
%\bibitem{Banerjee:2008wq}
%  R.~Banerjee, S.~Kulkarni,
%  ``Hawking Radiation, Covariant Boundary Conditions and Vacuum States,''
%  Phys.\ Rev.\  {\bf D79 } (2009)  084035.
%  [arXiv:0810.5683 [hep-th]].

%\cite{Gangopadhyay:2008ub}
%\bibitem{Gangopadhyay:2008ub}
%  S.~Gangopadhyay,
%  ``Anomalies, Horizons and Hawking radiation,''
%  Europhys.\ Lett.\  {\bf 85 } (2009)  10004.
%  [arXiv:0809.4572 [hep-th]].

%\cite{AlvarezGaume:1983ig}
\bibitem{AlvarezGaume:1983ig}
  L.~Alvarez-Gaume, E.~Witten,
  ``Gravitational Anomalies,''
  Nucl.\ Phys.\  {\bf B234 } (1984)  269.

%\cite{Bertlmann:2000da}
\bibitem{Bertlmann:2000da}
  R.~A.~Bertlmann, E.~Kohlprath,
  ``Two-dimensional gravitational anomalies, Schwinger terms and dispersion relations,''
  Annals Phys.\  {\bf 288 } (2001)  137-163.
  [hep-th/0011067].

%\cite{Minwalla:1999xi}
%\bibitem{Minwalla:1999xi}
%  S.~Minwalla, N.~Seiberg,
%  ``Comments on the IIA (NS)five-brane,''
%  JHEP {\bf 9906 } (1999)  007.
%  [hep-th/9904142].

%\cite{Narayan:2001dr}
%\bibitem{Narayan:2001dr}
%  K.~Narayan, M.~Rangamani,
%  ``Hot little string correlators: A View from supergravity,''
%  JHEP {\bf 0108 } (2001)  054.
%  [hep-th/0107111].

%\cite{Maldacena:1996ix}
%\bibitem{Maldacena:1996ix}
%  J.~M.~Maldacena, A.~Strominger,
%  ``Black hole grey body factors and d-brane spectroscopy,''
%  Phys.\ Rev.\  {\bf D55 } (1997)  861-870.
%  [hep-th/9609026].

%\cite{Das:1996wn}
%\bibitem{Das:1996wn}
%  S.~R.~Das, S.~D.~Mathur,
%  ``Comparing decay rates for black holes and D-branes,''
%  Nucl.\ Phys.\  {\bf B478 } (1996)  561-576.
%  [hep-th/9606185].

%\cite{Gubser:1996xe}
%\bibitem{Gubser:1996xe}
%  S.~S.~Gubser, I.~R.~Klebanov,
%  ``Emission of charged particles from four-dimensional and five-dimensional black holes,''
%  Nucl.\ Phys.\  {\bf B482 } (1996)  173-186.
%  [hep-th/9608108].

%\cite{Gubser:1996zp}
%\bibitem{Gubser:1996zp}
%  S.~S.~Gubser, I.~R.~Klebanov,
%  ``Four-dimensional grey body factors and the effective string,''
%  Phys.\ Rev.\ Lett.\  {\bf 77 } (1996)  4491-4494.
%  [hep-th/9609076].

%\cite{Klebanov:1996gy}
%\bibitem{Klebanov:1996gy}
%  I.~R.~Klebanov, M.~Krasnitz,
%  ``Fixed scalar gray body factors in five-dimensions and four-dimensions,''
%  Phys.\ Rev.\  {\bf D55 } (1997)  3250-3254.
%  [hep-th/9612051].

%\cite{Klebanov:1997cx}
%\bibitem{Klebanov:1997cx}
%  I.~R.~Klebanov, S.~D.~Mathur,
%  ``Black hole grey body factors and absorption of scalars by effective strings,''
%  Nucl.\ Phys.\  {\bf B500 } (1997)  115-132.
%  [hep-th/9701187].

%\cite{Zhang:2009jn}
\bibitem{Zhang:2009jn}
  B.~Zhang, Q.~-y.~Cai, L.~You {\it et al.},
  ``Hidden Messenger Revealed in Hawking Radiation: A Resolution to the Paradox of Black Hole Information Loss,''
  Phys.\ Lett.\  {\bf B675 } (2009)  98-101.
  [arXiv:0903.0893 [hep-th]].

%\cite{Zhang:2009td}
\bibitem{Zhang:2009td}
  B.~Zhang, Q.~-y.~Cai, M.~-s.~Zhan {\it et al.},
  ``Entropy is Conserved in Hawking Radiation as Tunneling: a Revisit of the Black Hole Information Loss Paradox,''
  Annals Phys.\  {\bf 326 } (2011)  350-363.
  [arXiv:0906.5033 [hep-th]].


\end{thebibliography}
\end{document}